\documentclass[12pt,preprint]{aastex}
\usepackage{epsfig}

\shorttitle{Universality in the distribution of caustics}
\shortauthors{Yano et al.}
\begin{document}
\title{Universality in the 
distribution of caustics \\in the expanding Universe}

\author{Taihei Yano$^1$, Hiroko Koyama$^1$, Thomas Buchert$^{1,2,3}$ 
and Naoteru Gouda$^1$}

\affil{$^{1}$
National Astronomical Observatory, 
Mitaka, Tokyo 181--8588, Japan\\
$^{2}$
Department of Physics and Research Center for the Early Universe, 
School of Science, The University of Tokyo, Bunkyo--ku, Tokyo 113--0033, Japan\\
$^{3}$
Theoretische Physik, Ludwig--Maximilians--Universit\"{a}t,\\
Theresienstrasse 37, D--80333 M\"unchen, Germany\\
E--mails: yano@pluto.mtk.nao.ac.jp, hiroko.koyama@nao.ac.jp,
buchert@theorie.physik.uni-muenchen.de, naoteru.gouda@nao.ac.jp}

\begin{abstract}
We numerically investigate the long--time evolution of density perturbations 
after the first appearance of caustics in an expanding cosmological model with
one--dimensional `single--wave' initial conditions.
Focussing on the time--intervals of caustic appearances
and the spatial distribution of caustics at subsequent times, we
find that the time--intervals of caustic appearances
approach a constant, i.e., their time--subsequent ratio converges to $1$;
it is also found that the spatial distribution of caustics at a given
time features some universality rules, e.g., the ratio between 
the position of the nearest caustic from the center and
that of the second nearest caustic from the center approaches a constant.
Furthermore we find some rules for the mass distribution for each caustic.
Using these universality constants we are in the position to predict the spatial 
distribution of caustics at an arbitrary time in order to give an estimate 
for the power spectral index in the fully--developed non--dissipative 
turbulent (`virialized') regime.
\end{abstract}

\noindent
{\it Subject headings}:
cosmology:theory -- large--scale structures -- self--similar -- non--dissipative
gravitational turbulence

\section{INTRODUCTION}

The evolution of large--scale structure in an expanding universe model is
one of the most important and interesting problems in cosmology.
It is generally believed that these structures have
formed owing to gravitational instability. After structure has formed,
the combined action of multi--stream forces and 
gravity is responsible for their stability in time 
(Buchert \& Dom\'\i nguez 1998), eventually leading to `virialized' structures
on certain spatial scales such as galaxies and clusters of galaxies.
Such equilibrium states, if they persist in time, are not necessarily 
in thermodynamical equilibrium, but are the consequence of a process that
was termed by Gurevich \& Zybin (see their review 1995 and refs. therein) 
{\it non--dissipative gravitational
turbulence}, emphasizing the role of multi--stream forces that counteract
the gravitational action and so eventually lead to equilibrium structures,
supported by velocity dispersion in collisionless systems.

Here we consider collisionless particles
as candidates for cold dark matter, and focus our interest on
universal behavior in the gravitational collapse process. 
The hope is to discover 
some rules that govern the multi--streaming hierarchy in a self--gravitating
system. In this research note we shall concentrate on the simplest model
of a plane--symmetric `single--wave' perturbation on an expanding 
universe model; the discovery of universality laws needs a detailed study at
high--spatial resolution. An advantage of plane symmetry is the 
existence of exact solutions that we shall incorporate in our numerical
scheme. Other aspects of multi--streaming have already 
been studied in three spatial dimensions (see, e.g., 
Kofman et al. 1994), and there is a plethora of works on powerlaw, 
fractal, or self--similar scaling properties of the density field, which are
all indirectly related to the understanding of multi--stream systems.
Also numerical simulations of self--gravitating systems 
practically evolve a multi--stream flow on smaller spatial scales.

Recently, Yano \& Gouda (1998) have studied the time evolution 
of density perturbations in the plane--symmetric system. They showed
that the power spectrum of the evolved perturbations can be
roughly separated into five regimes,
having initially power--law shape and a cutoff scale as shown in Figure 1.

According to these results, we can first separate the power spectrum roughly
into three regimes on the scales larger than the cutoff scale:
one is the linear regime ($k < k_{nl}$: Regime 1).
The value of the power index in this regime 
remains the initial power index, $n$.
The second regime is the single--caustic regime 
($k_{nl} < k < k_{snl}$: Regime 2).
In this regime, the power index becomes $-1$ and is independent of the initial 
conditions. This result is caused by the appearance of caustics at this scale, 
and these caustics determine the power index of 
the power spectrum in this regime.
The third regime is the multi--caustics regime 
($k_{snl} < k < k_{cut}$: Regime 3).
In this regime, caustics on various scales of waves determine
the power spectrum.
Therefore, on these scales, the power spectral index $\mu$ 
can be traced back to initial conditions.

On the scales smaller than the cutoff scale the problem is more involved. 
Nevertheless, we can roughly separate the power spectrum into 
two more regimes:
after the appearance of the first caustic, the power index in the regime 
of the scale smaller than the cutoff scale becomes $-1$.
This value is, as mentioned above, caused by the appearance of caustics.
However, thereafter, more and more caustics appear and so the separation of 
caustics becomes smaller and smaller.
Even after the appearance of many caustics, on scales smaller than the 
characteristic separation of the caustics ($k > k_{cs}$: Regime 5), 
the power index of the power spectrum can be obtained by the 
density profile around the singular point:
the power index on these scales also becomes $-1$, a result that can
be derived from catastrophe theory (or the Lagrange--singularity theory,
respectively), see, e.g., Arnol'd et al. 1982, Kotok \& Shandarin (1988), 
Gouda \& Nakamura (1988, 1989), Yano \& Gouda (1998). 
On the other hand, on the scales larger than the characteristic
separation of the caustics, in the fully developed `virialized' regime
($k_{cut} < k < k_{cs}$: Regime 4),
the distribution of the singular points determines the power index on
these scales instead of the density profile around one singularity:
we may say that the smoothed density profile with the smoothing
scale ($k_{cut}<k<k_{cs}$) determines the power index.
For a comprehensive description of the above discussion, see
Yano \& Gouda (1998).

In this paper we focus on the fully developed `virialized' multi--stream 
regime ($k_{cut} < k < k_{cs}$)
to clarify the physical process of the appearances of the caustics,
and we are going to estimate a value for the power index 
of the power spectrum in this regime.
For this purpose we study the time--intervals of the appearances of the caustics 
as well as the spatial distribution of the caustics at subsequent times.
The initial condition is chosen as `single--wave' perturbation
defined by $v=v_0\sin x$,
where $v$ is the velocity, $x$ is the position of a sheet,
and $v_0$ is an arbitrary constant. The evolution of such a  perturbation 
determines the power spectral index in the `virialized' regime, 
essentially Region 4 in Figure 1.
Accordingly we succeed to find some universal rules about the time 
when each caustic appears,
the spatial distribution of the caustics at subsequent times, and the
distribution of mass for each caustic.

Our paper is organized as follows.
In $\S 2$, we will briefly explain the numerical method to calculate
the time evolution of density perturbations in the expanding 
universe model.
In $\S 3$ we show the numerical results providing some rules for the
time--appearances of caustics, the spatial distribution of caustics at
subsequent times, and the mass distribution for each caustic. Here,
we also evaluate the power index of the power spectrum in the `virialized'
regime. The final section is devoted to a summary and a brief outlook.

\section{MODEL AND NUMERICAL METHOD}

We investigate the time evolution of a plane--symmetric density perturbation in
an Einstein--de Sitter universe model by using a semi--numerical method.
In the one--dimensional system many plane--parallel sheets move only 
in perpendicular direction to the surface of these sheets.
When two sheets cross, they are allowed to pass through eachother freely.
In this sheet system, there is an exact solution
until two sheets cross over as follows
(Zel'dovich 1970, Sunyaev \& Zel'dovich 1972, Doroshkevich et.al. 1973, 
Zentsova \& Chernin 1980, Buchert 1989, Yano \& Gouda 1998):
\begin{eqnarray}  
x &=& q + B_1(t)S_1(q) + B_2(t)S_2(q), \nonumber \\
v &=& \dot{B_1}(t)S_1(q) + \dot{B_2}(t)S_2(q)\;\;, 
\label{1}
\end{eqnarray}  
where $q$ and $x$ are the Lagrangian and the (comoving) Eulerian coordinates, 
respectively.
Here, $S_1(q)$, and $S_2(q)$ are arbitrary functions of $q$. 
$B_1(t)$, and $B_2(t)$ are the growing mode and the decaying mode of linear 
perturbation solutions, respectively.
We are considering the Einstein--de Sitter background universe model, i.e., 
$B_1(t)=a$ and $B_2(t)=a^{-\frac{3}{2}}$,
where $a$ is the scale factor. The velocity $v$ is the peculiar--velocity
normalized by the scale--factor.
We can compute the crossing time of all neighboring pairs of sheets.
We use the shortest of these crossing times as a time step.
Then, we can compute the new positions and velocities for all sheets 
at this crossing time.
After two sheets cross, we exchange the velocities of two sheets that
just crossed.
Then we obtain again $S_1(q), S_2(q)$, and therefore exact solutions
as follows:
\begin{eqnarray}  
S_1(q)=\frac{3}{5}a^{-1}(x-q)+\frac{2}{5}\dot{a}^{-1}v, \nonumber \\
S_2(q)=\frac{2}{5}a^{\frac{3}{2}}(x-q)
 -\frac{2}{5}\dot{a}^{-1}a^{\frac{5}{2}}v\;\;.
\end{eqnarray}  
These new exact solutions can be used until again two sheets cross.
In this way we obtain the exact loci of the sheets 
by coupling these solutions.
This semi--numerical method has good accuracy, because 
we connect exact solutions.

In our calculation we are going to use $2^{12}$ sheets.
Periodic boundary conditions are fixed at a length of $2\pi$. 
Then we consider the evolution of the `single--wave' perturbation
whose wave number is $1$. 

\section{RESULTS}

In this section we calculate the time--intervals for the appearances of
caustics, the spatial distribution of caustics at subsequent times, and
the mass distribution for each caustic. Accordingly we find some rules
for the evolution of the caustic distribution. 
Using these rules, we are going to estimate the
power index of the power spectrum in the `virialized' regime.

\subsection{Universality rules and constants for the occurence of caustics}

We examine the time evolution of the `single--wave' perturbation
setting $S_1(q) =\sin q$ and $S_2(q) =0$ in Equations (\ref{1}).
Then, comoving positions $x$ and scaled peculiar--velocities $v$ of
each sheet are simplified to: 
\begin{eqnarray}  
x &=& q + B_1(t)\sin(q)\;\; , \nonumber \\
v &=& \dot{B_1}(t)\sin(q)\;\;,
\end{eqnarray}
see Figure 2-a.
Hereafter we call this case the 
{\it single--wave case}.

The first caustic has just appeared at the center of the x--axis ($x=0$)
in Figure 2-b. After the first appearance of a caustic, more caustics
appear at the center (phase mixing starts in phase space) (Figures 2-c -- 2-d)
(Doroshkevich et. al. 1980; Melott 1983; Gouda \& Nakamura 1989; Yano \& Gouda 
1998). According to this cascade of appearing caustics, 
the density distribution 
evolves as in Figures (3-a -- 3-d).
Figures 2-a, 2-b, 2-c, and 2-d correspond to Figures 3-a, 3-b, 3-c, and 3-d, 
respectively.
As we see from Figures 3-a, 3-b, 3-c, and 3-d, a hierarchy of nested density
peaks appears in the course of time.

In order to discover rules for the distribution of the caustics at
an arbitrary time,
we sort the time--intervals in the appearances of the caustics, 
the spatial distribution of the caustics,
and the mass distribution for each caustic at a certain time 
as follows.

First, we examine the time--intervals related to the appearances of the caustics.
The solid line in Figure 4 shows the time when caustics appear
at the center of the x--axis (a so--called A3--type singularity).
The dashed line shows the time--intervals for the appearances of caustics.
As we can see from Figure 4, these time--intervals are almost constant.
In order to clarify this point, we introduce the ratio of those
intervals defined by $T_i \equiv (t_{i+2}-t_{i+1})/(t_{i+1}-t_i)$,
where $t_i$ is the time when the i--th caustics appears.
Then, the relation between $T_i$ and the time when the $i$--th caustic appears
is plotted in Figure 5.
We see a loss of accuracy after around the 10th caustic appeared.
Nevertheless, independent of the number of sheets,
we see that the value $T_i$ quickly converges to $1$.
We can understand why the ratio becomes 1 as follows:
according to the time evolution of the density distribution,
more and more caustics appear at the center of the x--axis and these
caustics may be viewed as forming a bound cluster.
After the evolution of this cluster decouples from the expansion of the 
universe model, sheets around the center of the cluster behave like a
harmonic oscillator due to the fact that an enhanced gravitational
potential forces particles with velocities less than the escape velocity
from the cluster to rebound. The periods 
of the oscillation of the sheets finally becomes independent of the amplitude.
Therefore, it is reasonable that the time--intervals of the appearances of
caustics becomes constant, i.e., the value of  $T_i$ becomes 1.

Second, we consider the spatial distribution of caustics.
In Figure 6 we plot the spatial positions of caustics when the 
N--th caustic appears at the center of the x--axis.
Here we note that the number of appearances of
caustics can be used instead of time, because we have already shown that the
intervals may be considered practically constant.
We define the position of a caustic, $X(i,j)$, as the $i$--th 
position of a caustic at the time when the $j$--th caustic appears. 
Then, for example, 
$X(i,i)$ represents the position of the $i$--th caustic 
that has just appeared as A3--type at the center of the x--axis.
Now, we examine the  spatial positions of caustics at a certain time ($N=N_0$).
We adopt $N_0=10$ in our analysis because the accuracy of the positions 
of caustics decreases for $N_0 > 10$ as shown in Figure 5.
The solid and dashed lines in Figure 7 denote the spatial
position of the $N$--th caustic, $X(N,10)$, and the spatial interval
between the (N-1)--th and the N--th caustics, $X(N-1,10)-X(N,10)$, respectively.
Furthermore we depict the corresponding ratios:
the solid and dashed lines in Figure 8 depict
the ratio of spatial positions $X(N,10)/X(N-1,10)$
and the ratio of spatial intervals
$(X(N-1,10)-X(N,10))/(X(N-2,10)-X(N-1,10))$, respectively.
In Figure 8 we see no constant in the spatial positions or 
spatial intervals of caustics at first glance.

However, we see a constant when we plot positions of the nearest
caustic from the center, $X(N-1,N)$, or the ratio $X(N-1,N)/X(N-2,N-1)$.
In Figure 9, the relation between 
the position $X(N-1,N)$ and the time when the N--th caustic appears is plotted.
The ratio of positions $X(N-1,N)/X(N-2,N-1)$ is plotted in Figure 10.
In Figure 10, $R_1$ represents the ratio between
the position of the nearest caustic from the center and
that of the second nearest caustic from the center $X(N-1,N)/X(N-2,N)$.
$R_2$ represents the ratio between the nearest caustic from the
center at the (N-1)th appearance of a caustic and that at the Nth appearance of
a caustic $X(N-1,N)/X(N-2,N-1)$.  Both ratios feature constants:
the ratio $R_1$ and $R_2$ have values of around $0.15$ and $0.3$, respectively.

If we plot positions of caustics by using a 
logarithmic scale of spatial positions and a linear scale of time, 
we can see some ``lattice'' as shown in Figure 11. This apparent 
regularity is the result of the existence of constants in the ratios.
Here we note that calculated positions are not exact for smaller
than about $10^{-6}$ because of finite sheet resolution.
Therefore, the ``lattice'' is destroyed after around the 10--th appearance of 
caustics.


Third, we investigate the mass distribution for each caustic at a certain time
$(N=N_0)$.
Here, we define the mass, $M(i,N)$, as proportional to the number of sheets 
which are located inbetween $X(i,N)$ and $X(i+1,N)$ at the time, when the Nth 
caustic appears. The schematic
picture of $M$ is shown in Figure 12.

Then, we obtain the mass distribution for each caustic at $N_0=10$, which
is shown in Table \ref{tbl-1} and Figure 13.
We deduce the following relation between $i$ and $M$ for large values
of $i$:
\begin{equation}
M(i,N)\propto R_M^i, \qquad\qquad {\rm where} \qquad R_M\sim 0.5\;\;.
\end{equation}
Since we know the spatial distribution of the caustics and the mass
distribution for the caustics, we can construct the smoothed density profile 
to calculate the power index of the power spectrum in the `virialized' regime.


\subsection{Evaluation of the power spectrum in the `virialized' regime}

In this subsection we estimate the index of the power spectrum in the
`virialized' regime, using the rules which we have found in the
previous subsection.
Assuming that the spatial distribution of caustics obeys the law of 
a geometric series,
\begin{equation}
 X(i,N_0)=a_0a(N_0)^i \;\;,
\end{equation}
that is,
\begin{equation}
i=\frac{\ln X(i,N_0)-\ln a_0}{\ln a}\;\;,
\end{equation}
and also that the mass distribution for each caustic obeys the law of 
another geometric series,
\begin{equation}
 M(i,N_0)=b_0b(N_0)^i\;\;,
\end{equation}
we find the density profile as:
\begin{eqnarray}
\label{density-profile}
\varrho(X)&=&\frac{M(i,N_0)}{X(i,N_0)-X(i+1,N_0)}
=\frac{b_0}{(1-a)a_0}\left(\frac{b}{a}\right)^{-\ln a_0/\ln a}
X^{\frac{\ln (b/a)}{\ln a}}\;\;.
\end{eqnarray}
Its Fourier transform gives 
\begin{eqnarray}
\delta_k&\propto& \int \varrho(X)e^{ikX}dX\nonumber\\
&\propto& \int X^{\frac{\ln (b/a)}{\ln a}}e^{ikX}dX\nonumber\\
&\propto&k^{-1-\frac{\ln (b/a)}{\ln a}}\;\;.
\end{eqnarray}
Finally we obtain the power spectrum as:
\begin{eqnarray}
P(k)\equiv |\delta_k|^2 &\propto& k^{-2-2\frac{\ln (b/a)}{\ln a}}\;\;.
\end{eqnarray}
In fact, we have found  $a=R_1=0.15$ and $b=R_M=0.5$ from the previous 
subsection, so that we obtain the power spectrum in the explicit form: 
\begin{eqnarray}
\label{final}
P(k) \propto k^{\nu}, \qquad \qquad {\rm where} \qquad \nu=-0.73.
\end{eqnarray}
On scales smaller than the characteristic scale $k>k_{cs}$ (Regime 5 in
Figure 1), the power index of the power spectrum becomes $-1$. This
is because the density profile around one singular point determines 
the power spectrum on this scale.

On scales larger than the characteristic scale $k<k_{cs}$, 
on the other hand, the density profile in the spatial interval,
where the singular points lie closer than the scale of this interval, 
is smoothed, and the power index of the power spectrum
deviates from $-1$.
The larger scales we consider, the more  
the density profile is smoothed.
The power index in the `virialized' regime $k_{cut}<k<k_{cs}$,
accordingly, is determined by
the smoothed density profile (\ref{density-profile}), and therefore
the result for the power index of the power spectrum 
under the smoothed density profile is given by (\ref{final}).

Our result shows that the power index in the `virialized' regime 
$k_{cut}<k<k_{cs}$
may be approximated by the index value $\nu=-0.73$.
In addition, let us note that the wavenumbers $k_{cs}$, which represent the
characteristic separations of caustics, becomes larger and larger 
as time increases, since the characteristic separation of caustics
becomes shorter and shorter.


\section{SUMMARY AND OUTLOOK}

We have investigated the evolution of density perturbations
after the first appearance of a caustic, exemplified for a plane--symmetric
`single--wave' perturbation in an expanding universe model.
Following Yano \& Gouda (1998), who showed that the power spectrum can be
roughly separated into five regimes with different power--law index,
we have explored the fourth regime. 
The detailed analysis of a many--sheet system is needed in this regime in order
to determine the time--evolution and the asymptotics of the power spectral 
index.

We have argued that the system of caustics displays some universality,
after some time, with respect to regularity of time--intervals 
in the appearances of caustics, and the spatial distribution of caustics 
at subsequent times.

First, 
we have noticed the existence of a rule for the time--intervals 
of the appearances of caustics: 
the ratio of intervals quickly approaches $1$, a fact that we have made
plausible.  
Second, 
we have found some rules hidden in the spatial distribution of caustics:
the ratio between 
a position of the nearest caustic from the center and
that of the second nearest caustic from the center $R_1 =X(N-1,N)/X(N-2,N)$
approaches a value of around $0.15$.
Furthermore,
the ratio between the nearest caustic from the
center at the (N-1)th appearance of a caustic and that at the Nth appearance of
a caustic $R_2= X(N-1,N)/X(N-2,N-1)$ attains a value of around $0.3$.

Third, we have found a rule for the mass distribution for each caustic,
that is, we have derived the density distribution of a cluster.
The mass distribution for each caustic can be fit to a line with the geometric
progression of ratio $0.5$. 

Furthermore, using the above constants, 
we have found the smoothed density
profile, and finally have evaluated the power index of the power spectrum 
in the `virialized' regime to be $-0.73$.
This value is different from $-1$ that is determined by the density
profile around one singular point.
In other words, the power spectum in the region below
the cutoff scale can be divided into two clearly distinct regions, 
the `single--caustic' regime and the `virialized' regime.
These results are consistent with the numerical results 
by Yano \& Gouda (1998).

The discovery presented in this research note, namely  
that there exist remarkable rules for the distribution
of caustics, points towards several interesting theoretical implications
for long--term properties of self--gravitating systems.
As in studies of dynamical systems we here deal with a hierarchy of 
bifurcations. A fruitful way to explore universality properties further is
to use renormalization group techniques similar to those developed for 
the Feigenbaum scenario of pitchfork bifurcations (see: Feigenbaum 1978, 1979;
Derrida et al. 1979; Collet \& Eckmann 1980). 
With this note we have just provided a hint to the existence of universal
constants. A detailed renormalization group and fixpoint analysis is needed
to support these findings. In this context an interesting question has to 
remain open: the existence of chaotic regimes. 

Besides this theoretical interest we think that our results may be useful
to check the accuracy of numerical simulations concerning their 
stability under long--time integration.

In this paper we have focused on one non--trivial regime,
the `virialized' regime (Regime 4 in Figure 1), and have succeeded to evaluate 
the power index of the power spectrum in this regime semi--numerically.
In order to complete the understanding of the evolution of the power spectra 
of density fluctuations, we have also to understand the behavior of the power index
in the multi--wavenumbers regime (Regime 3 in Figure 1).

Furthermore, we could employ the systematic framework of the velocity 
moment hierarchy to eventually derive analytically universal behavior of
the density profile; also, nonlinear statistical methods beyond
the power spectrum would give more refined information.

\acknowledgments

{\small This work was supported in part by 
a Grant--in--Aid for Scientific Reserach (B) (No. 15340066).
TB acknowledges support by the National Observatory Tokyo where this work was
started, as well as hospitality at the University of Tokyo with
support by the Research Center for the Early Universe (RESCEU, Tokyo),
during which discussions on this work were continued. 
Also, partial support by the Deutsche Forschungsgemeinschaft 
(within SFB 375) is acknowledged.}


\begin{figure}
\plotone{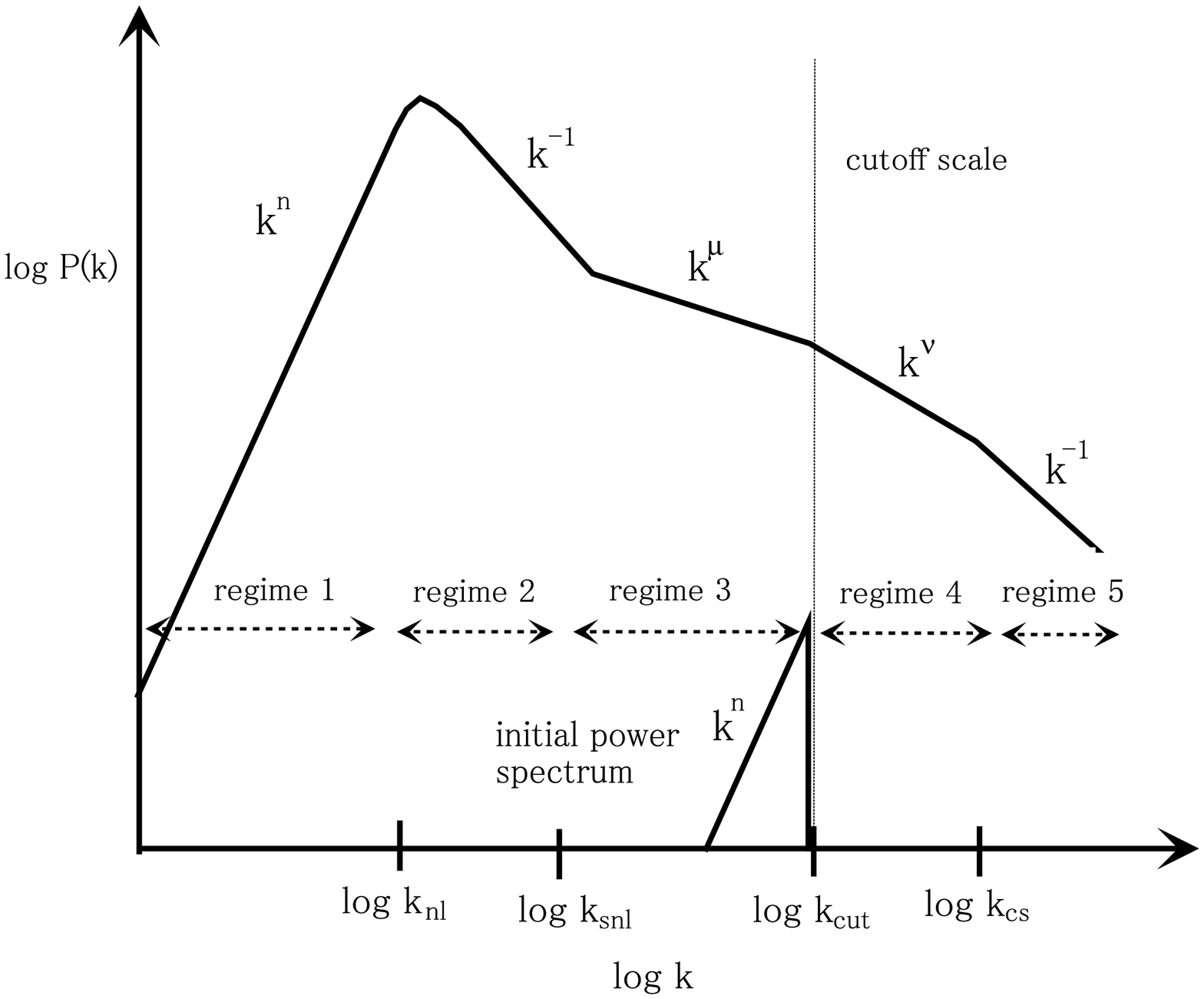}
 \caption{Schematic representation of the evolved power spectrum for a general 
initial power spectrum after the first appearance of caustics.\label{fig1}
}  
\end{figure}

\begin{figure}
\plottwo{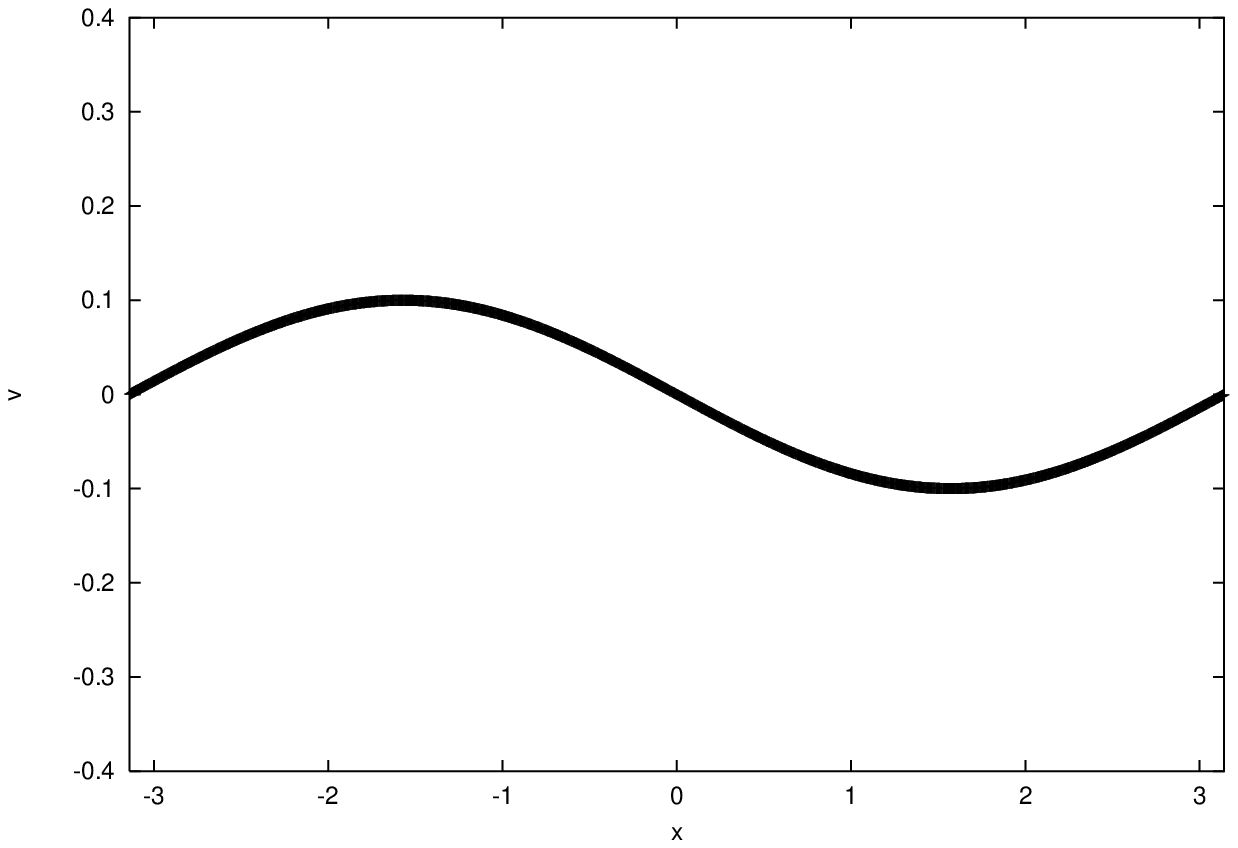}{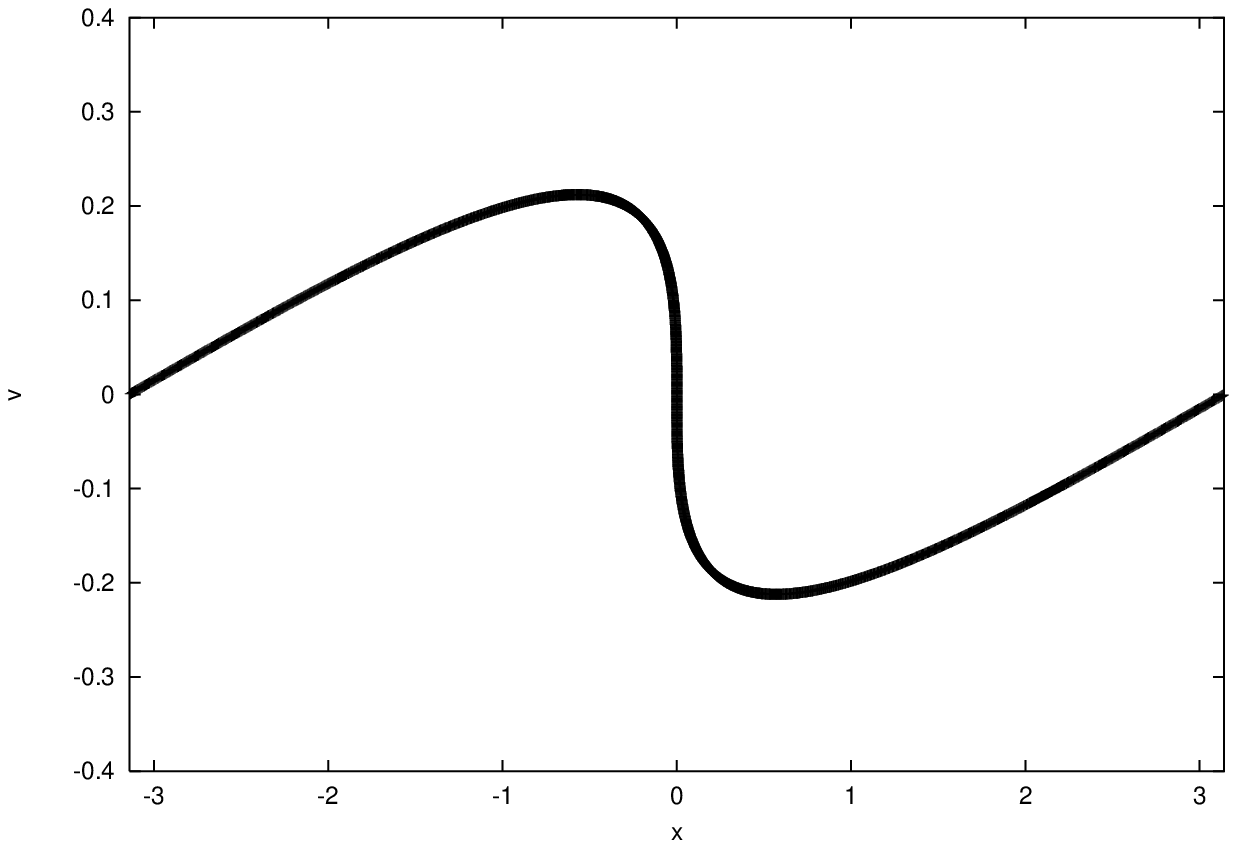}

\plotone{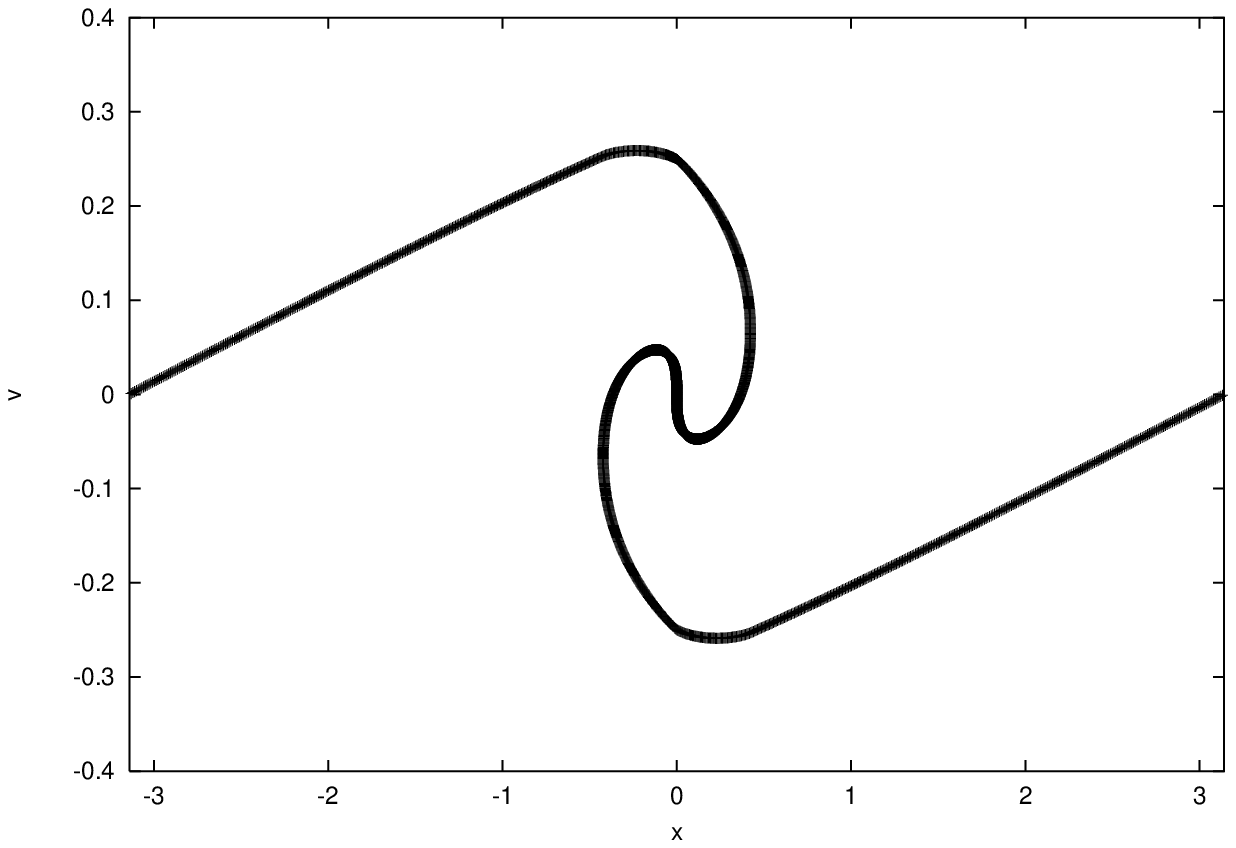}
\plotone{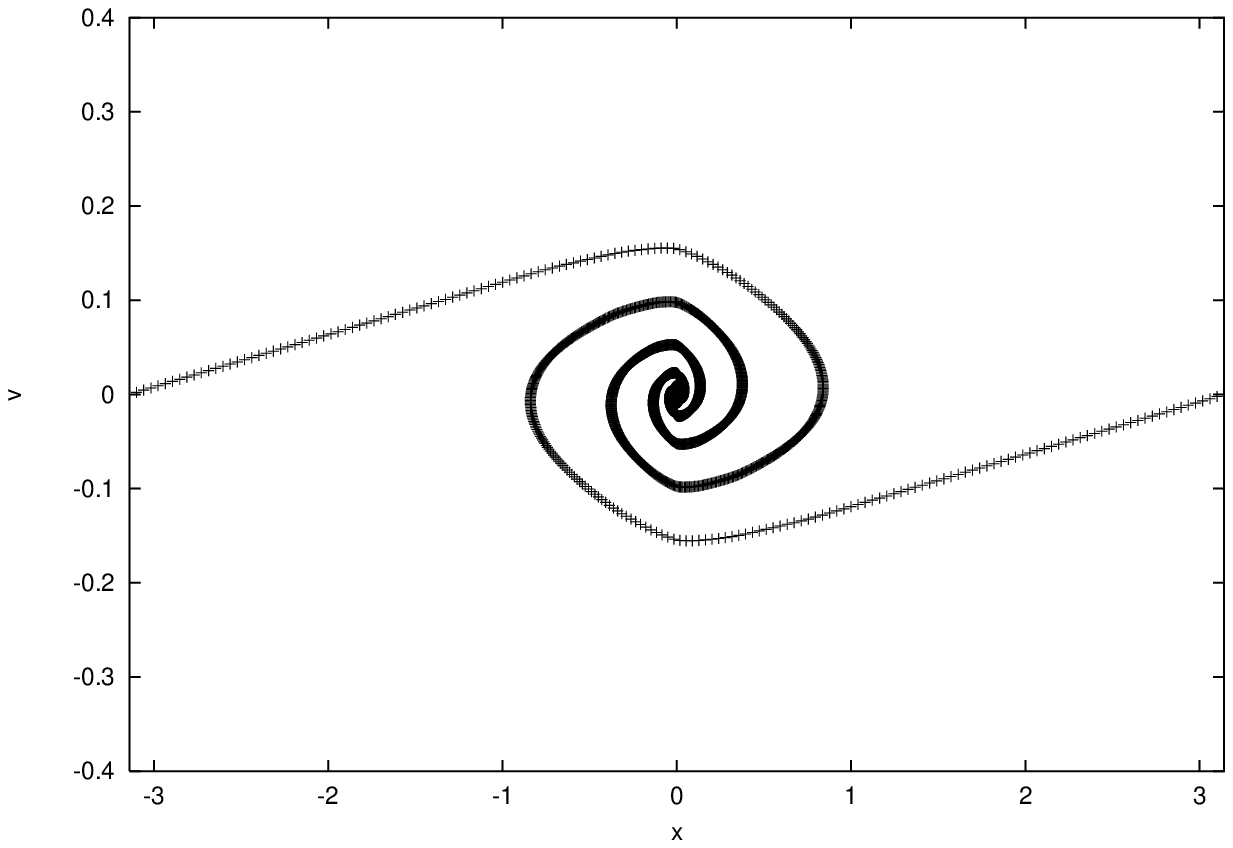}
 \caption{(a) Distribution in phase space for the `single--wave' case
at the initial time (a), at the first appearance of caustics (b),
at the second appearance of caustics (c), and 
at the fifth appearance of caustics (d).\label{fig2}
}  
\end{figure}

\begin{figure}
\plottwo{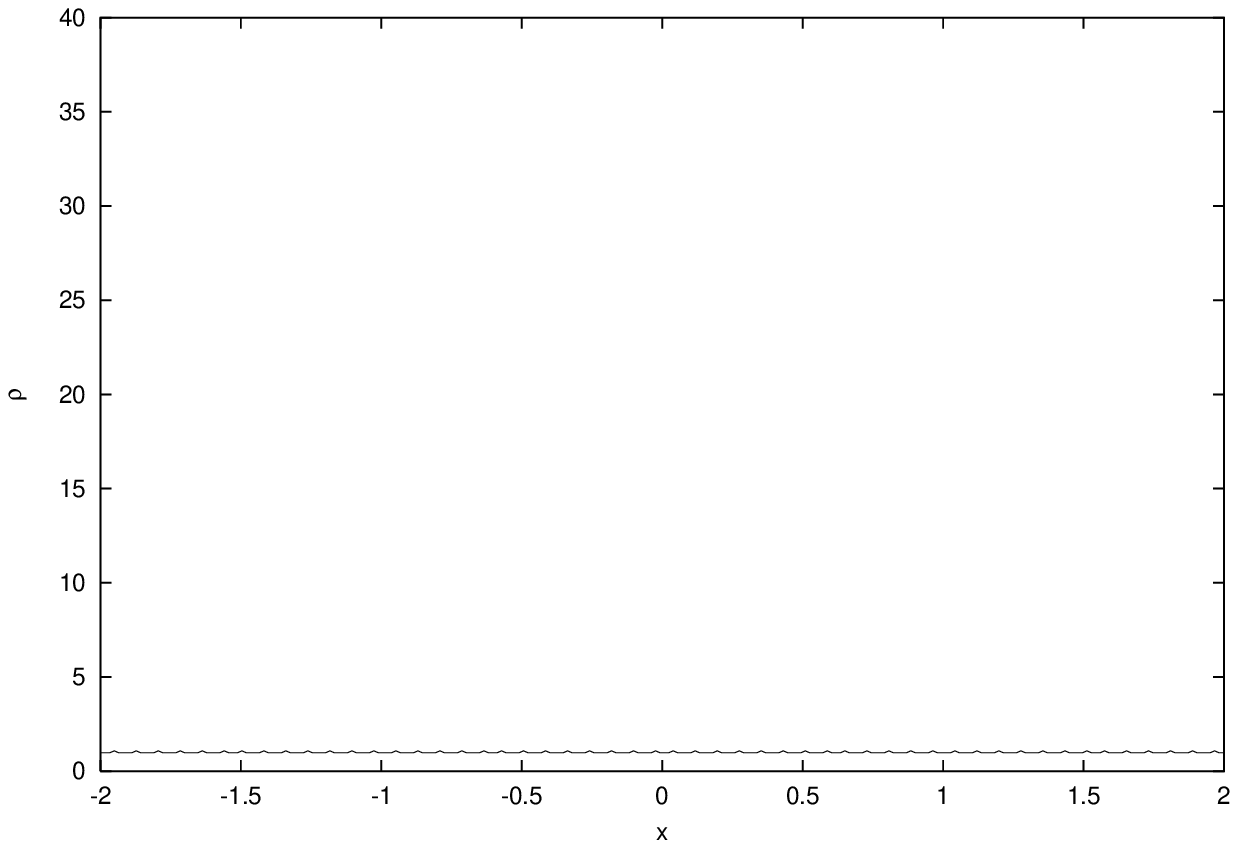}{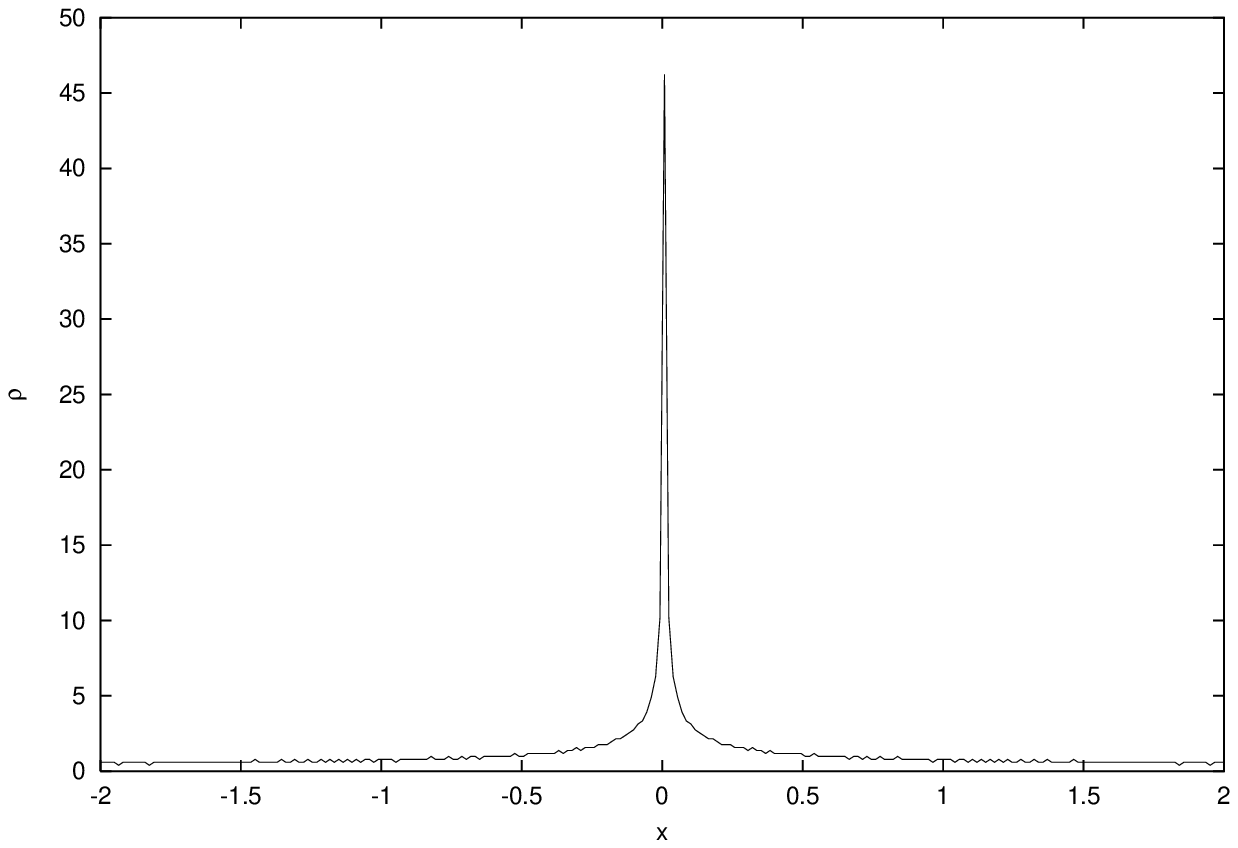}

\plotone{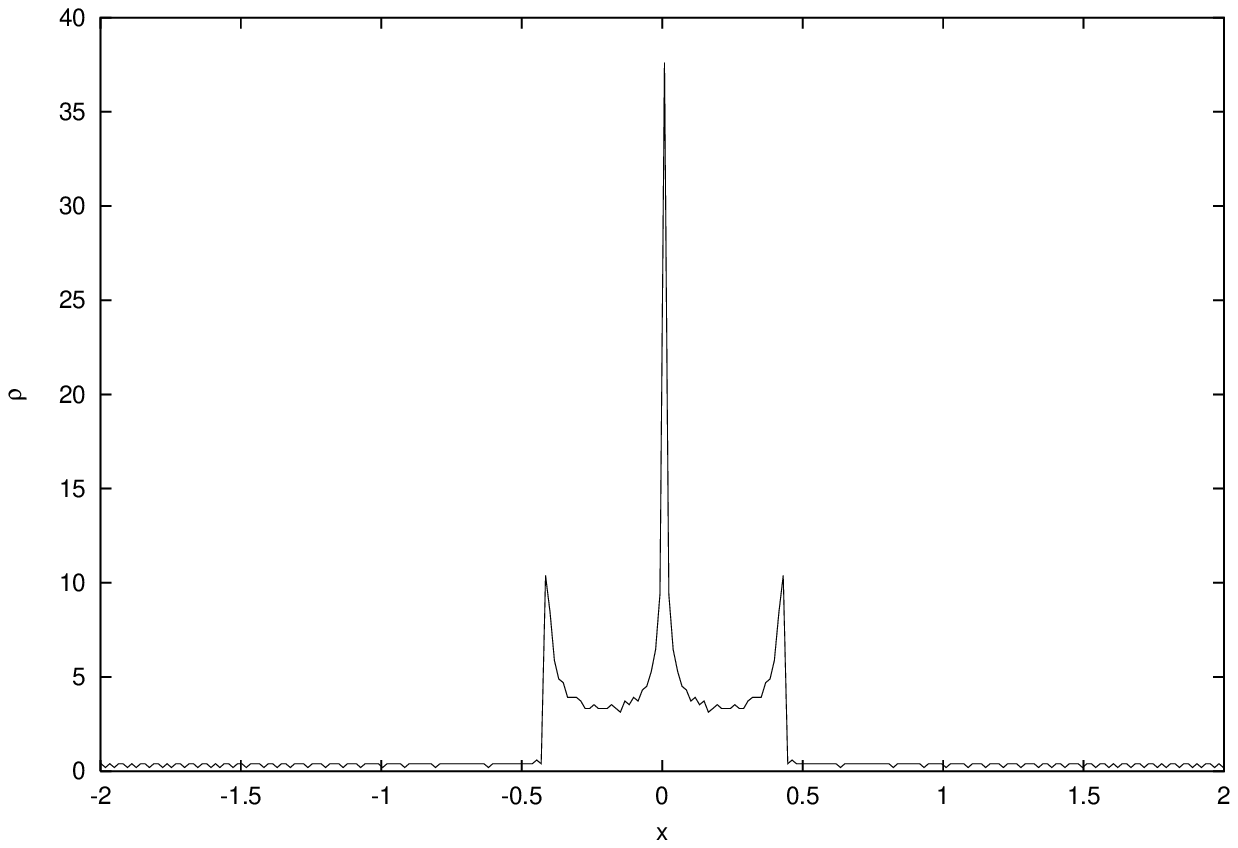}
\plotone{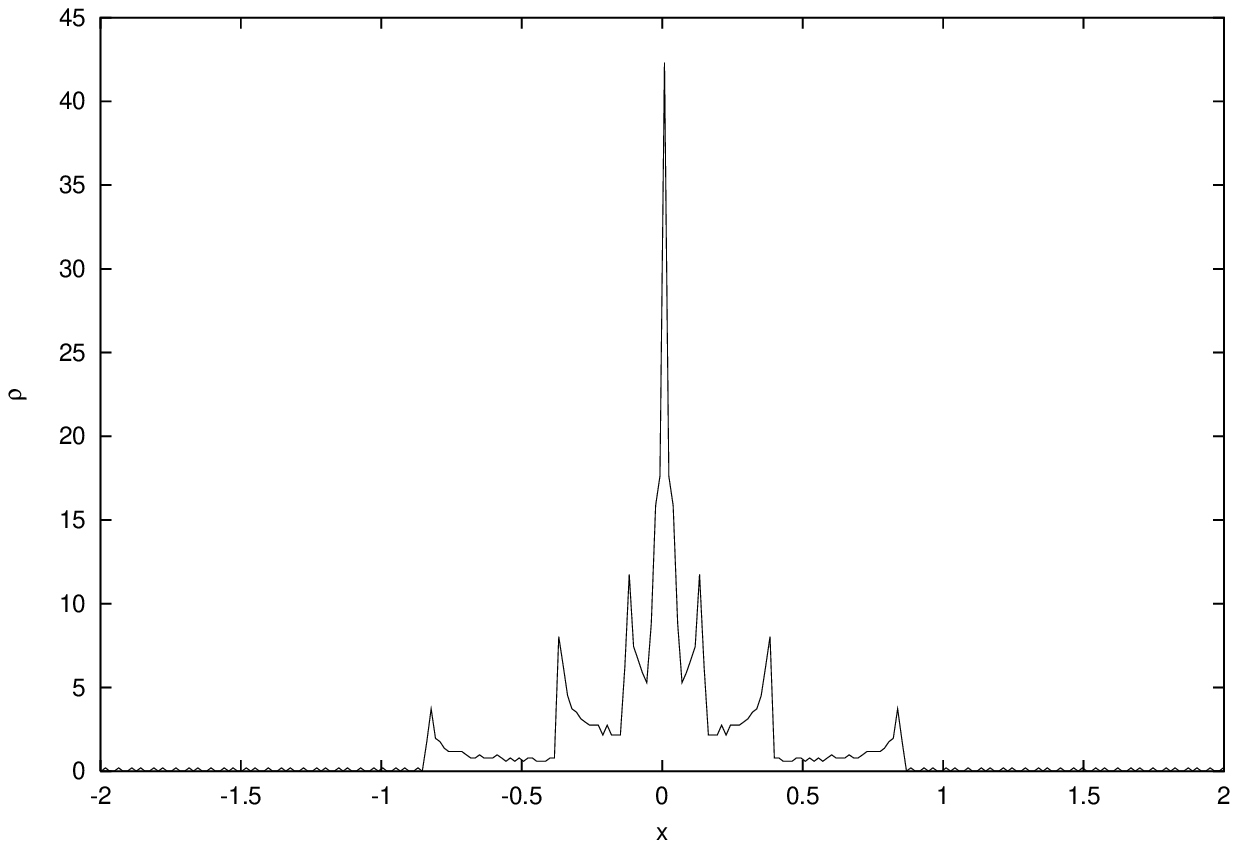}
 \caption{ (a)Density distribution for the `single--wave' case 
at initial time (a), and at the first (b), the second (c) and the
fifth (d) appearance of caustics. \label{fig3}
}  
\end{figure}

\begin{figure}
\plotone{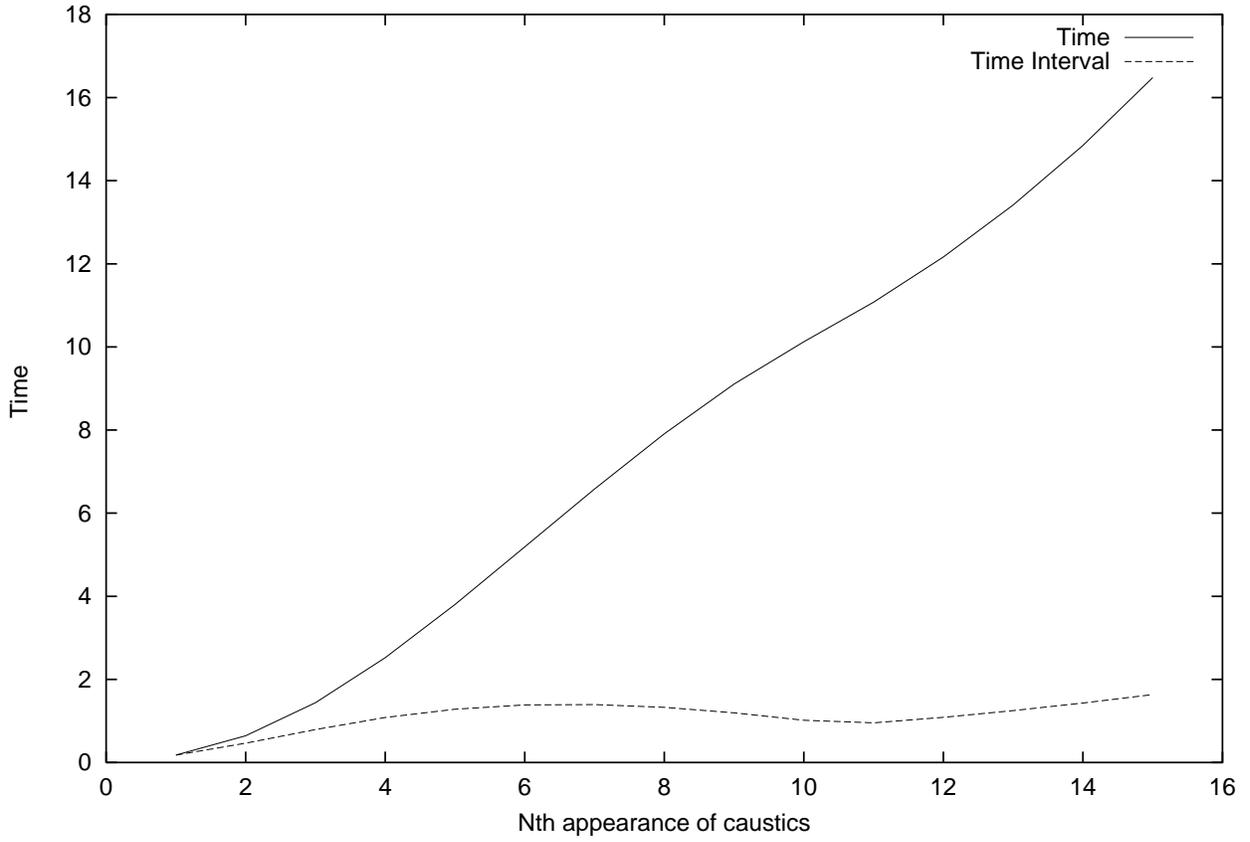}
 \caption{Times and intervals for the appearance of caustics at the center of 
the x--axis. The solid and dotted lines represent times and intervals, 
respectively. \label{fig4}
}  
\end{figure}

\begin{figure}
\plotone{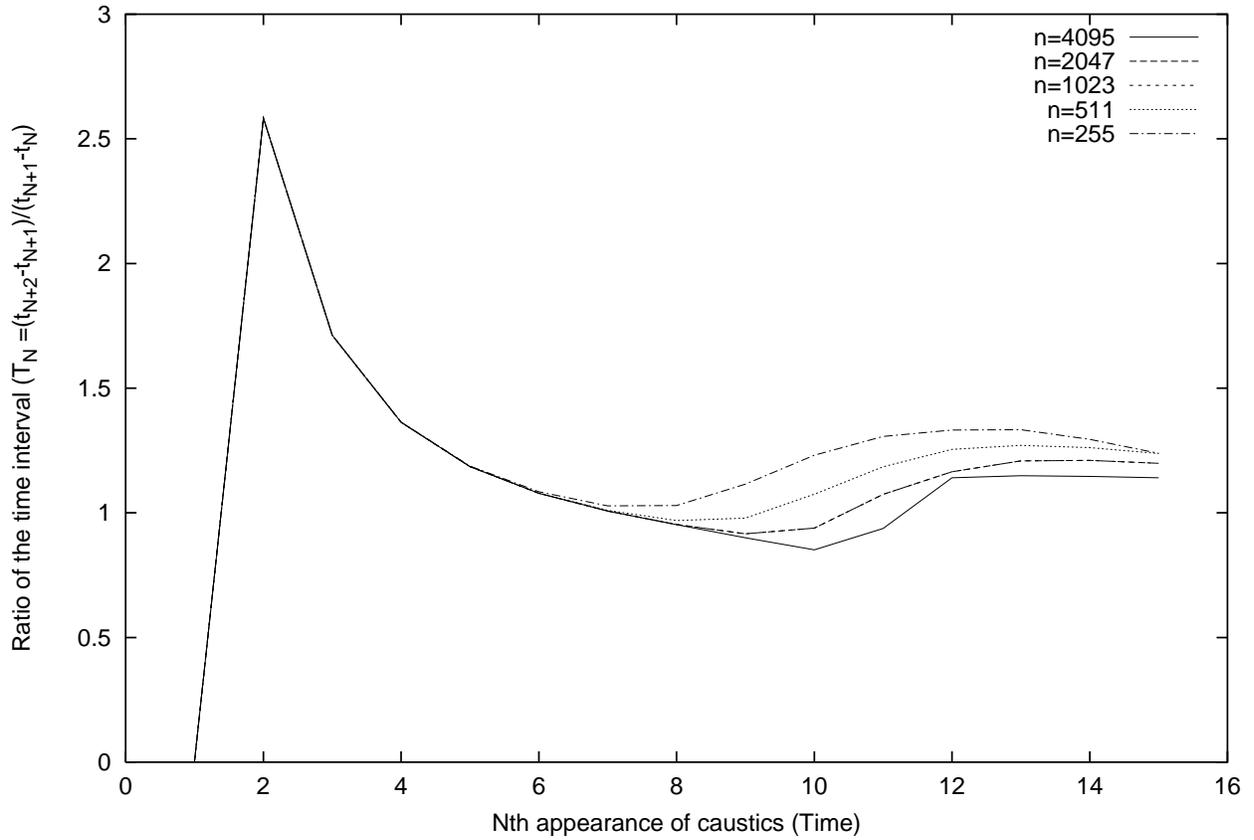}
 \caption{Ratio of intervals in the appearance of caustics. 
The solid, long--dashed, short--dashed, dotted, and dash--dotted lines
are cases for different numbers of sheets: 
n=4095, n=2047, n=1023, n=511, and n=255, respectively.\label{fig5}
}  
\end{figure}

\begin{figure}
\plotone{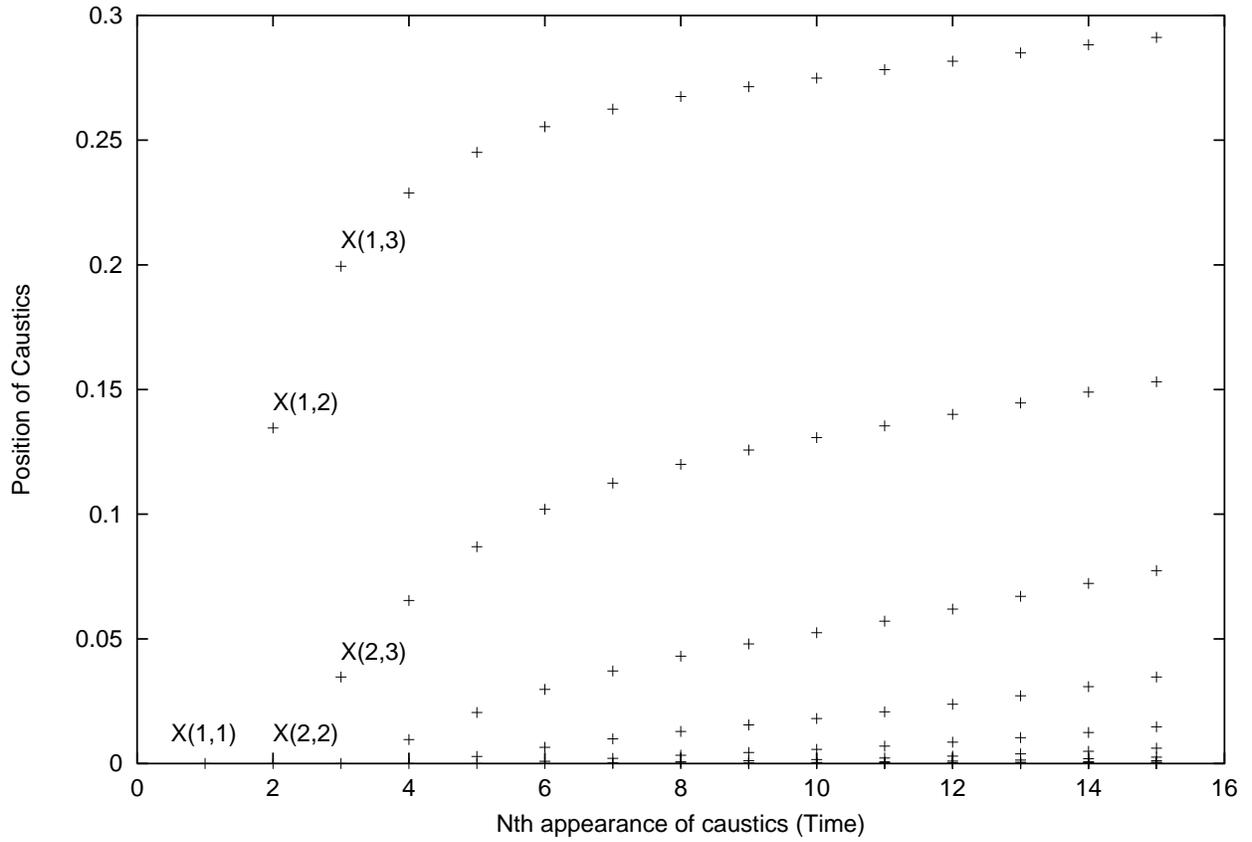}
 \caption{Spatial distribution of caustics at the N--th appearance 
of caustics (the {\it bifurcation diagram}).\label{fig6}
}  
\end{figure}

\begin{figure}
\plotone{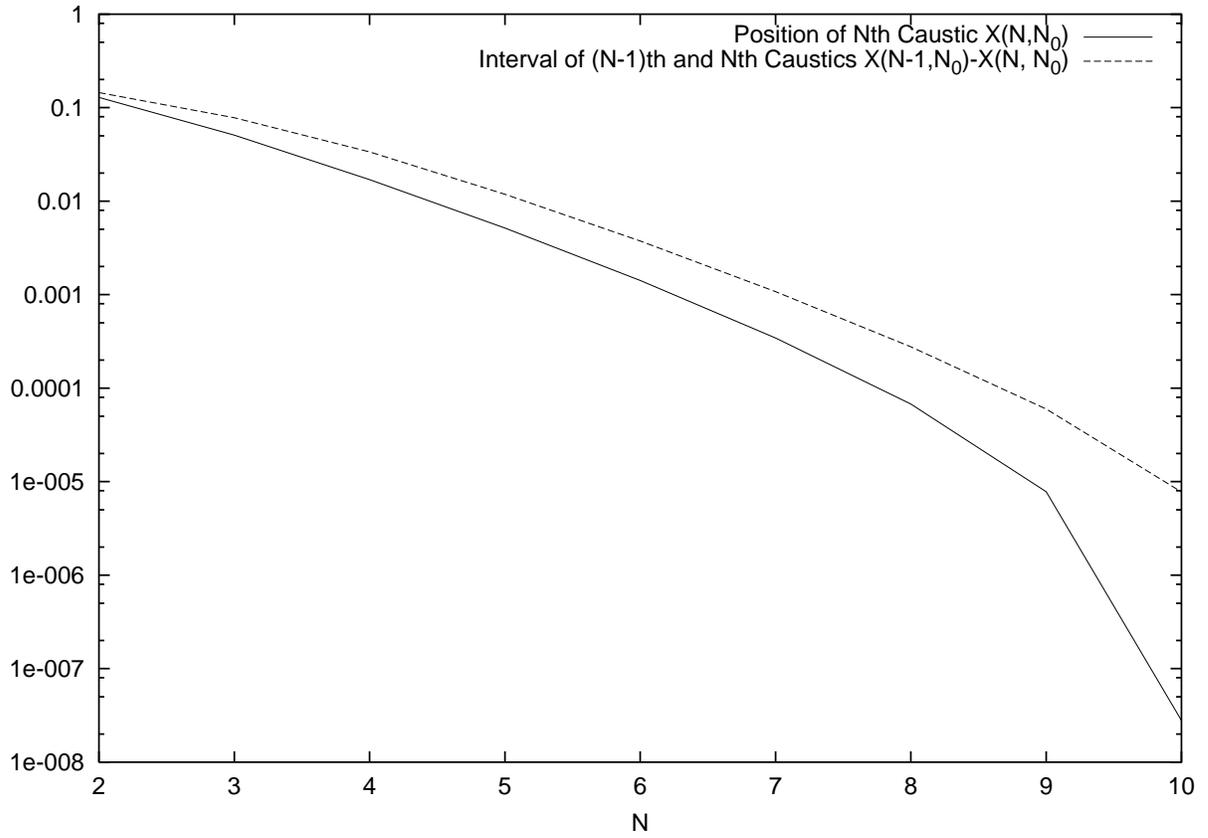}
 \caption{Spatial positions and intervals of caustics at the 10--th appearance 
of caustics. The solid and dashed lines represent the spatial positions
and intervals, respectively. Smaller N represents a position that is further
away from the center.\label{fig7}
}  
\end{figure}

\begin{figure}
\plotone{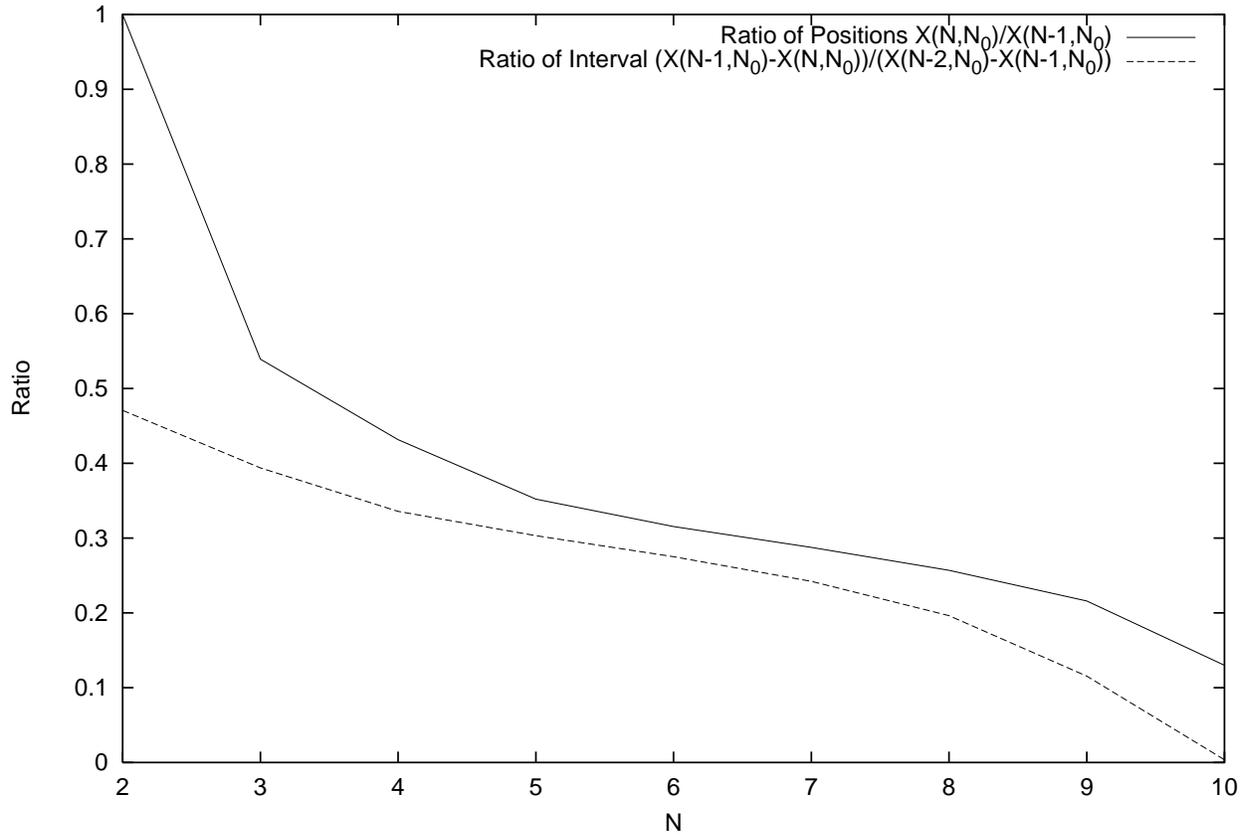}
 \caption{Ratio of spatial positions and intervals 
of caustics at the 10--th appearance 
of caustics. The solid and dashed lines denote ratios of spatial positions
and intervals, respectively.\label{fig8}
}  
\end{figure}

\begin{figure}
\plotone{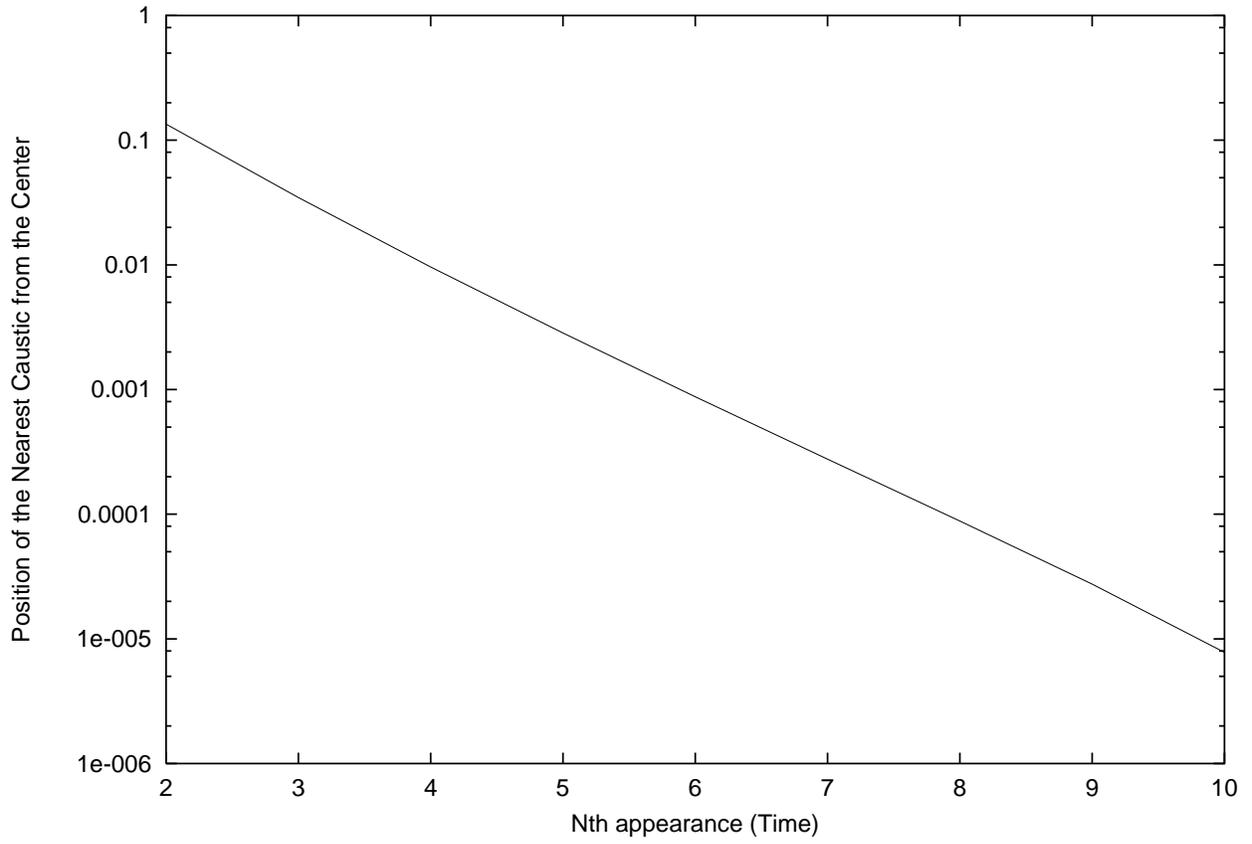}
 \caption{Spatial position of the caustic that is nearest to 
the center of the x--axis $X(N-1,N)$.\label{fig9}
}  
\end{figure}

\begin{figure}
\plotone{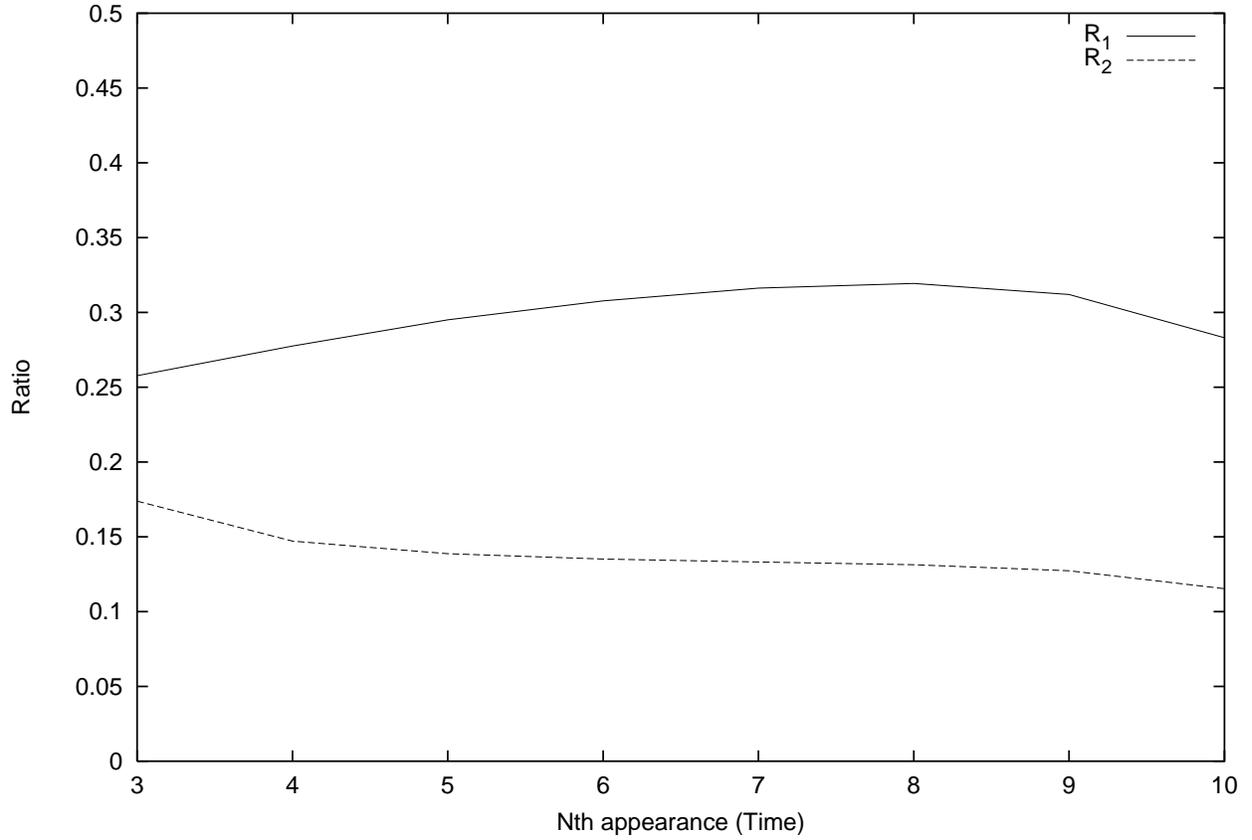}
 \caption{Ratio of positions of caustics.
$R_1$ represents the ratio between
the position of the nearest caustic from the center and
that of the second nearest caustic from the center $X(N-1,N)/X(N-2,N)$.
$R_2$ represents the ratio between the nearest caustic from the
center at the (N-1)th and that at the Nth caustic appearance,
$X(N-1,N)/X(N-2,N-1)$.\label{fig10}
}  
\end{figure}

\begin{figure}
\plotone{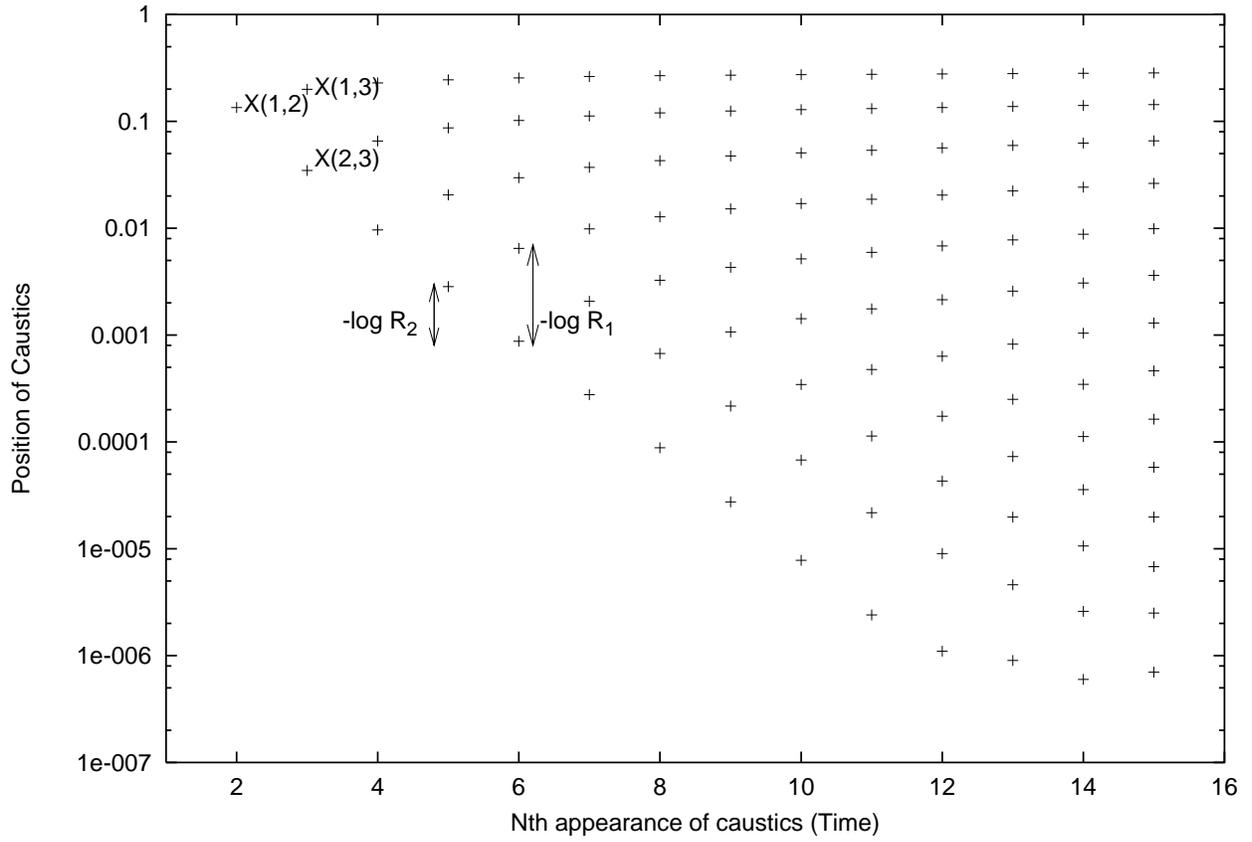}
 \caption{Spatial positions of caustics at the N--th appearance of caustics.\label{fig11}
}  
\end{figure}

\begin{figure}
\plotone{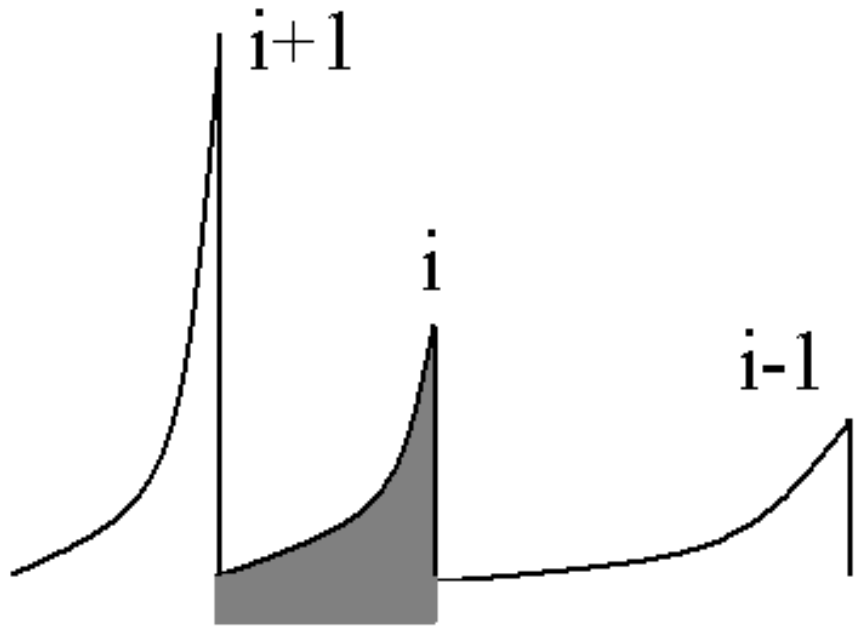}
\caption{Schematic picture of the density distribution and 
the definition of the mass for the i--th caustic (the shaded region).\label{fig12}}  
\end{figure}

\begin{figure}
\plotone{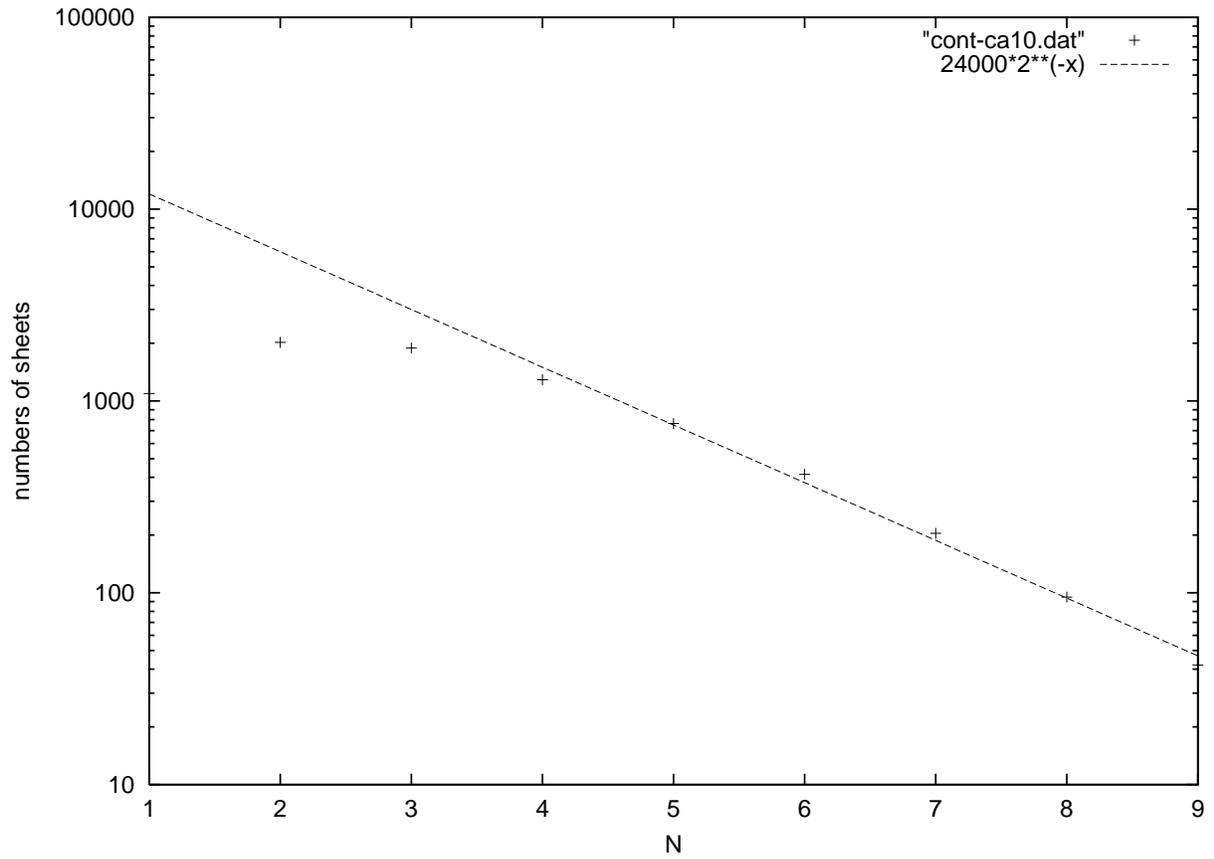}
 \caption{Mass distribution contributing to each caustic at the 10--th
 appearance of caustics.\label{fig13}}  
\end{figure}

\begin{table}
\begin{center}  
\caption{Mass distribution for each caustic when the 10th caustics appear.}  
\label{tbl-1}
\begin{tabular}{ccc}
\hline
\hline
order of caustics&position of caustics&numbers of sheets\\
\hline
1& 0.2736580 &1094\\
2& 0.1288644 &2021\\
3&0.0507813&1888\\
4&0.0170624&1289\\
5&0.0051794&763\\
6&0.0014272&414\\
7&0.0003467&204\\
8&0.0000683&95\\
9&0.0000081&42\\
\hline
\end{tabular}
\end{center}
\end{table}

\end{document}